\documentclass[12pt]{article}
\usepackage{amsmath}
\usepackage{latexsym}
\usepackage{bbm}
\usepackage{amssymb}

\def\L{\Lambda}

\def\and{{\rm and}}

\def\ie{{\it i.e.,} }

\def\and{{\quad {\rm and} \quad}}

\def\A5{{AdS_5 \times S^5}}

\newcommand{\be}{\begin{equation}}
\newcommand{\bea}{\begin{eqnarray}}

\newcommand{\ee}{\end{equation}}
\newcommand{\eea}{\end{eqnarray}}

\begin{document}
\vspace*{-1.0in}
\thispagestyle{empty}
\begin{flushright}
CALT-TH-2024-013
\end{flushright}

\normalsize
\baselineskip = 18pt
\parskip = 6pt

\vspace{1.0in}

{\Large \begin{center}
{\bf Comments Concerning a Hypothetical \\
Mesoscopic Dark Dimension}
\end{center}}

\vspace{.25in}

\begin{center}
John H. Schwarz\footnote{jhs@theory.caltech.edu}
\\
\emph{Walter Burke Institute for Theoretical Physics\\
California Institute of Technology\\ Pasadena, CA  91125, USA}
\end{center}

\vspace{.25in}

\begin{center}
\textbf{Abstract}
\end{center}
\begin{quotation}
Motivated by string-theoretic swampland conjectures, the existence of
a dark fifth dimension, whose size is roughly 1 -- 10 microns, has been proposed.
A great deal of supporting evidence has been presented, and definitive
experimental tests are likely to be carried out.
The basic idea is that the four-dimensional spacetime that we
observe is confined to a brane that is localized in
the dark dimension. This short note points out that there are two distinct
ways to realize such a scenario in string theory/M-theory. In the one considered previously,
the dark dimension is topologically
a circle and our observable 4d spacetime is confined to a brane that
is localized in a GUT-scale region of the circle.
An alternative possibility is that the dark dimension
is a line interval with ``end-of-the-world-branes" at each end.
The latter option would imply the existence of a parallel 4d spacetime,
with the same gravitational forces as ours, but otherwise its own physical laws,
microns away from us!
\end{quotation}

\newpage


\pagenumbering{arabic}


\newpage


Motivated by string-theoretic swampland conjectures \cite{Agmon:2022thq},
it has recently been proposed that there is a {\em mesoscopic} fifth dimension
whose radius $\rho$ is roughly 1 -- 10 microns \cite{Vafa:2024fpx}--\cite{Montero:2022prj}.
One aspect of this is the proposal that a tower of light states in our universe
is related to the smallness of the cosmological constant $\Lambda$ \cite{Lust:2019zwm}.
The new idea is that this tower corresponds to Kaluza--Klein (KK) modes of a
graviton in a dark dimension, and that they constitute dark matter. The size
of this dimension is of order $\Lambda^{-1/4}$.
(We always set $\hbar =c=1$.) The unknown coefficient is responsible
for the uncertainty in the prediction for $\rho$.
In this way dark energy and dark matter are connected without invoking an anthropic principle.
This dimension is called {\em dark}, because it only supports gravitational forces.
The basic idea is that the four-dimensional spacetime that we
observe lives on a {\em brane} that is localized at a point
(more precisely, a GUT-scale region) in
the dark dimension (DD). Any additional compact dimensions must be very much
smaller than the DD. Thus, for many
purposes, we may assume that the brane is well approximated by a standard-model
3-brane (SM3-brane) even though it would be desirable to understand exactly how such
a brane arises from a more fundamental string theory/M-theory starting point.
For reasons explained in \cite{Vafa:2024fpx}, the existence of more than one
mesoscopic dimension is excluded. As discussed in \cite{Obied:2023clp},
astrophysical constraints limit the allowed size of the DD to 1 -- 10 microns,
and future large-scale structure surveys could support or rule out its existence.

A number of arguments were made in \cite{Vafa:2024fpx}--\cite{Montero:2022prj},
based on various theoretical and observational facts, in support of the DD proposal.
These arguments suggest that the size of the DD is not very far below the
limit (about 30 microns) obtained by a recent experimental study of
Newton's $1/r^2$ gravitational force law at very short distances \cite{Lee:2020zjt}.
That experiment used ``a stationary torsion-balance detector and a rotating
attractor containing test bodies." There have also been other related
experiments \cite{Geraci:2008hb}--\cite{Blakemore:2021zna}.
An experiment that can probe even shorter distances is being planned \cite{Baeza-Ballesteros:2021tha}\cite{Baeza-Ballesteros:2023par}. It will
study the orbits of a microscopic planetary system. Precession of the orbit
would imply a central potential different from $1/r$. Other experimental
approaches may be possible, e.g. based on atom interferometry.

It is now conceivable that the DD realization of string theory
will receive experimental support within this decade.
Prior to the DD proposal I did not think that string theory could make novel
testable predictions in the foreseeable future.
This DD proposal appears to be well motivated, but even so it should be considered
speculative. The experiments ought to be a very high priority, because
experimental verification would be revolutionary.
A negative result would not be revolutionary, but it would still be
fundamental and very important.

There have been previous theoretical proposals for a relatively large extra
dimension, such as \cite{Antoniadis:1998ig} and \cite{Randall:1999ee}.
They differ in important respects from the
current proposal, which is motivated by swampland
considerations as well as various astrophysical arguments. In particular,
the current proposal shows promise for giving a deeper understanding
of why the amounts of dark matter and dark energy are comparable
without invoking the anthropic principle.

If a mesoscopic DD is discovered, it will become important to figure out how
it fits into string theory. We have known since the mid-1990s that the
various versions of superstring theory/M-theory are related by dualities, so that
there really is a unique underlying theory. However, the different versions,
which focus on different corners of the moduli space of vacua, can appear
more or less promising as starting points for phenomenology. There have
been two such classes of starting points that account for almost all studies
of string phenomenology. One class is based on the $E_8 \times E_8$ heterotic
string \cite{Gross:1984dd}.
The other class is based on F-theory \cite{Vafa:1996xn}, which utilizes
certain nonperturbative solutions of type IIB superstring theory.
See \cite{Marchesano:2024gul} for a recent discussion of the status
of string phenomenology.

An F-theory approach to realizing the DD proposal is described in \cite{Montero:2022prj}.
In that realization the DD is topologically a circle, and the SM3-brane is localized
in a GUT-scale region of the circle.
Among other things, that paper argues that this picture gives a tower of states,
associated to a 7-brane, at $10^9$ -- $10^{10}$ GeV. This scale, called the {\em species scale},
corresponds to the 5d Planck scale.


Following the discovery that cancellation of local gauge and
gravitational anomalies in 10d superstring theories
is only possible for the gauge groups $SO(32)$ and $E_8 \times E_8$ \cite{Green:1984sg},
a string theory with the latter gauge group, called the $E_8 \times E_8$ heterotic string,
was constructed \cite{Gross:1984dd}. For the following decade (prior to the introduction
of F-theory) this was the only version of superstring theory that showed promise for
giving realistic phenomenology. The introduction of
Calabi--Yau compactification \cite{Candelas:1985en}
launched a great deal of interest in superstring theory in general,
and this specific approach
in particular. It was clear from the outset that this
setup has many qualitative successes.
However, even today there is no specific construction that looks exactly right.
The math is hard, and the number of possibilities is very large.
The same can be said for the F-theory approach.

Among the many exciting discoveries made in the mid-1990s
was the realization that the $E_8 \times E_8$ heterotic string theory and the type IIA
superstring theory actually have an eleventh dimension
that is not visible in perturbation theory \cite{Horava:1995qa}\cite{Horava:1996ma}.
In the type IIA case there is a circular eleventh dimension ($S^1$)
whose size is proportional to the string coupling constant.
In the limit of infinite coupling
one obtains 11-dimensional M-theory. In the case of the $E_8 \times E_8$ theory there is an
orbifold eleventh dimension ($S^1/Z_2$), whose size also is proportional to the string coupling
constant. This dimension can  also be viewed as a line interval with two ends. Each of the
ends is attached to an ``end-of-the-world" 9-brane (an EW9-brane). One $E_8$ gauge field
is associated to each of the two EW9-branes.
This is required by the anomaly analysis \cite{Horava:1995qa}.
The 11-dimensional bulk, in between the EW9-branes, is described by 11-dimensional M-theory,
which is the UV completion of 11d supergravity \cite{Cremmer:1978km}.

The eleventh dimension of the $E_8 \times E_8$ theory is
a DD. Most studies of Calabi--Yau compactification of this theory have
ignored the eleventh dimension, which may be a reasonable thing to do for many purposes.
However, it is an important fact that this class of string theory solutions necessarily
implies the existence of a DD. Some of the implications of the DD have been discussed in
\cite{Witten:1996mz} -- \cite{Lukas:1998tt}. As explained in \cite{Witten:1996mz},
one consequence of increasing the string coupling constant (and hence the size of the DD)
is to reduce the lower bound on the 4d Newton
constant $G_N$. This is important, since $G_N$ is necessarily too large in the weakly coupled
$E_8 \times E_8$ theory. So the requirement that the DD in this theory must be quite large
has been understood for a long time. What is new now
is the suggestion, based on swampland considerations, that the size of this dimension
should be several microns. By the standards of particle physics, this is very large indeed.

Suppose we consider implementing the DD proposal using the $E_8 \times E_8$ theory.
Then an essential difference from the F-theory example is that it requires the
existence of {\em two} EW-branes. After a suitable compactification of six dimensions
they can be approximated by 3-branes, and one of them can
be identified as the standard model 3-brane (SM3-brane)
that describes our 4d reality at least up to energies of
a TeV and possibly much further. In this setup
there is a second parallel 3-brane, just microns away!
Should we expect the physics of the second EW-brane to be the same as or different from
``our" EW-brane? The Calabi--Yau phenomenology works best when
the two $E_8$'s are treated differently. This implies that the
physics of the two EW-branes is different, so that seems likely to be the case.
In any case, the possibility of a parallel 4d spacetime just microns away from us
is challenging to visualise and seems quite incredible! However, as the
F-theory example illustrates, this is not an
unavoidable prediction of the DD proposal -- just one of the two options.

For strong coupling the $E_8 \times E_8$ heterotic theory is not a string theory.
(A similar remark can be made for the type IIA theory.) In M theory
the basic branes are an M2-brane and its magnetic dual,
which is an M5-brane. A cylindrical
M2-brane can be stretched across the dark dimension with its
boundaries attached to the EW-branes. In the weak-coupling limit
the two EW-branes approach one another, and
the cylinder collapses to become the $E_8 \times E_8$ string. However,
if the branes are microns apart, it looks
nothing like a string -- unless it can be approximated by an open string
stretched between the two EW-branes, but I am unaware of any such proposal.
In the $E_8 \times E_8$ theory the
coupling constant $g$ is proportional to the radius of the dark dimension, $\rho$.
More specifically, $g \sim \rho/l_s$, where $l_s$ is the string scale,
which is very small. If $\rho$ is several microns, this ratio is enormous.
The problem is evident from the shape of the stretched M2-brane.
If there were an S-dual formulation of the $E_8 \times E_8$ heterotic string theory,
whose coupling constant is $\tilde g = 1/g$, that could be helpful.
The $SO(32)$ heterotic string theory has an S-dual formulation, namely the type I
superstring theory, which is based on a non-BPS (breakable) string. It seems unlikely
that there is an analogous S-dual of the $E_8 \times E_8$ heterotic string theory.
One problem is that it is not possible to define a $g \to \infty$ limit while
retaining both of the EW-branes, and we don't want $\rho$ to be infinite.

Suppose that experimental evidence for the existence of a DD is obtained
by  measuring the strength of gravity at short distances.
How could we decide whether the DD is a circle or a line segment
with a second brane at the other end of the DD?
We would only be sensitive to the gravitational effects of the other brane.
One could imagine a star on our brane orbiting an invisible
object that resides on the other brane. Would it be possible to establish that
the invisible object is not a black hole or something on our
brane that we already understand?

It is conceivable that a very sensitive test of gravity
at short distances could distinguish between the two geometries. If the DD is a circle
of radius $\rho = 1/\mu$, then summing Yukawa potentials for the KK excitations
of a 5d graviton gives a potential proportional to
\be
\frac{1}{r} \sum_{n=-\infty}^{\infty} e^{-|n| \mu  r} =
\frac{1}{r} \frac{1+e^{-\mu  r}}{1-e^{-\mu  r}} \sim \frac{1}{r} (1+ 2e^{-\mu  r} +\ldots)
\ee
In the case of the orbifold/line segment, one has instead
\be
\frac{1}{r} \sum_{n=0}^{\infty} e^{-|n| \mu  r} =
\frac{1}{r} \frac{1}{1-e^{-\mu  r}} \sim \frac{1}{r} (1+ e^{-\mu  r} +\ldots)
\ee
The leading large $r$ asymptotic behavior is shown. For small $r$ it is proportional
to $\rho/r^2$ in both cases, as expected, though they differ by a factor of 2.

These formulas probably require corrections for the following reason.
A massless graviton in 5d has five polarizations,
forming an irreducible representation of the SU(2) little group, as does
a massive spin two particle in 4d. The only mismatch here concerns the zero modes.
In both cases there are three extra massless modes in 4d with helicities $+1$, $0$, $-1$
in addition to the desired graviton with helicities $+2$ and $-2$.
They are not wanted, since it is well established that 4d gravity is mediated by
a massless spin 2 graviton. A plausible resolution of this problem is that
these three modes combine to give a massive spin one particle in 4d.
The mass of this vector particle must go to zero in the
decompactification limit, if one is to recover 11d M theory.
Therefore, it ought to be proportional to $\mu$.
The formulas above need to be modified in order to
correctly account for this massive vector. In principle,
the experiments should be able to reveal one massive
spin one particle in addition to the usual graviton and the
infinite tower of massive spin two particles
by measuring the strength of gravity at short distances
to high precision.

An analogous problem occurs when M theory has
a circular dimension giving strongly coupled type IIA superstring theory.
The type IIA massless spectrum in 10d contains
a graviton, a vector, and a scalar that descend from the
11d graviton. The presence of the massless vector reflects the
$U(1)$ isometry of the eleventh dimension, in contrast to the
cases we have discussed in which one or two branes break the symmetry.
In the type IIA case there is a known deformation of the theory in which the scalar and vector
combine to give a massive vector \cite{Romans:1985tz}.
This mass arises when one or more 8-branes are added
to the type IIA background \cite{Bergshoeff:1996ui}. Thus, it is clear that the vector
must be massive whenever branes break the $U(1)$ symmetry of the dark dimension.
This is the case for the F-theory and M-theory examples under consideration here.

The original papers \cite{Vafa:2024fpx}--\cite{Montero:2022prj} make a strong case for
KK excitations of gravitons in the DD comprising the most
important component of dark matter. Of course, it is still possible that
there are also other sources of dark matter.
Other sources of dark matter in the DD might include
KK excitations of sterile neutrinos or other bulk degrees of freedom.
Possible sources of dark matter on the SM3-brane
are the familiar ones that have received the most attention
by both theorists and experimentalists so far. These include axions and wimps,
though the latter have been largely excluded.
If there is a second EW-brane, matter that is confined to that brane would
also be dark matter.

All known black holes are much larger than the proposed DD.
Black holes that are smaller than the DD would be localized
within that dimension, away from any branes, and they would be very difficult to
detect. The implications of the DD proposal for primordial black holes have been
discussed in \cite{Anchordoqui:2022txe}-\cite{Anchordoqui:2024akj}. These papers
point out that in the DD scenario 5d black holes have a much longer lifetime
due to Hawking decay than 4d ones of the same mass. Therefore,
they suggest that primordial black holes that are smaller than the
DD could be another significant component of dark matter.

One may wonder whether the DD proposal has any bearing on the $\L$CDM model
of cosmology. In particular, the time evolution of various quantities is important.
Since $G_N \sim G_5/\rho$, where $G_5$ is the 5d Newton constant, and $\L \sim \rho^{-4}$,
it follows that
\be
\frac{\dot{G_N}}{G_N} = \frac{\dot{G_5}}{G_5} - \frac{\dot{\rho}}{\rho}
\and \frac{\dot{\L}}{\L} = - 4 \frac{\dot{\rho}}{\rho} .
\ee
Variation of $G_N$ is constrained by various observations, such as binary
pulsars, and variation of $G_5$ seems unlikely in the epoch
where the $\L$CDM model is supposed to apply. Therefore
significant variation of $\rho$ and $\L$ appears to be unlikely
in this approach. Even so, the DD model may help to
address the current stress in the $\L$CDM model. An additional element is the fact
that the DD model predicts that Kaluza--Klein excitations of the  graviton
constitute a spectrum of decaying dark matter particles. While there have been studies of
of decaying dark matter as a possible mechanism to address the stress, the details of how it
occurs in the DD model are different from the way it has been modeled in these studies.
There may also be other effects, such as Casimir forces, that are relevant.

In conclusion, while the existence of a dark dimension of the sort discussed
here is certainly speculative, it is a well-motivated and experimentally testable
proposal. It is based on plausible
string theory considerations  that should be experimentally testable. Its
confirmation would be revolutionary for particle physics and astrophysics.
If the existence of a DD is confirmed, an important question would be
whether or not there is a parallel 3-brane microns away from the SM3-brane on
which we reside.

\section*{Acknowledgments}

I am grateful to Cumrun Vafa and Temple He for helpful discussions.
This material is based upon work supported by the U.S. Department of Energy,
Office of Science, Office of High Energy Physics, under Award Number DE-SC0011632.

\end{document}